\documentstyle[aps,prl,multicol,psfig]{revtex}
\begin{document}
\draft
           
\title{Crossover behavior for long reptating polymers}
\author{Enrico Carlon$^1$, Andrzej Drzewi\'nski$^{2,3}$ and 
J.M.J. van Leeuwen$^2$}
\address{$^1$INFM - Dipartimento di Fisica, Universit\`a di Padova,
I-35131 Padova, Italy \\
$^{2}$Instituut-Lorentz, University of Leiden, P.O.Box 9506, 
2300 RA Leiden, the Netherlands \\
$^3$Institute of Low Temperature and Structure Research, 
Polish Academy of Sciences,
P.O.Box 1410, 50-950 Wroc\l aw 2, Poland }
\date{\today}
\maketitle

\begin{abstract}
We analyze the Rubinstein-Duke model for polymer reptation by means of density 
matrix renormalization techniques. We find a crossover behavior for a series 
of quantities as function of the polymer length. The crossover length may 
become very large if the mobility of end groups is small compared to that of 
the internal reptons. Our results offer an explanation to a controversy 
between theory, experiments and simulations on the leading and subleading
scaling behavior of the polymer renewal time and diffusion constant.
\end{abstract}

\begin{multicols}{2}\narrowtext

The study of the dynamical properties of polymers is a field of great
interest, because of important applications ranging from material science 
to biophysics.
The process of reptation, i.e. the motion of a polymer along its own 
contour by the diffusion of stored length, is generally believed to be 
one of the most important mechanisms for polymer dynamics \cite{deGennes,doi}. 
The simplest model for reptation is that introduced by Rubinstein \cite{Rubin} 
and later extended by Duke \cite{Duke} to include the effect of a driving 
field. In spite of its simplicity the Rubinstein--Duke (RD) model contains the
essential physics of reptation and compares well with experiments
\cite{experiments}. As very little exact results are available for the 
model one has to rely on numerical techniques to investigate its properties.

In this Letter we present results on the scaling behavior of the polymer 
renewal time and diffusion constant, which are obtained by the
Density--Matrix Renormalization--Group (DMRG) technique \cite{DMRGbook}. 
DMRG allows to compute stationary state properties for rather long polymers 
with unprecedented accuracy. 
We find a crossover
behavior with two different regimes: for an intermediate range of lengths 
various quantities scale as a function of the polymer length with 
{\it effective} exponents that differ from the asymptotic ones. We also show 
that by tuning an appropriate parameter it is possible to increase or 
decrease considerably the crossover length, i.e. the characteristic length 
which separates the two regimes.
The findings help to clarify controversial results between theory, 
simulations and experiments for the renewal time and diffusion constant.
Crossover behavior for single polymer reptation has been suggested earlier 
on various occasions \cite{crossover,Rubin}. Our results are an articulation 
and proof of these suggestions. Moreover they enable us to pinpoint the 
crossover region precisely and to estimate the effective exponents for 
arbitrarily long polymers. 

\begin{figure}[b]
\centerline{
\psfig{file=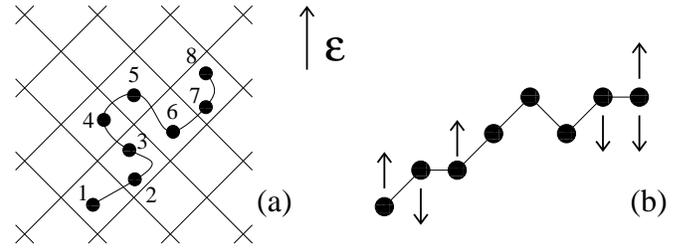,height=3.2cm}}
\vskip 0.2truecm
\caption{
(a) Configuration of a chain with $N=8$ reptons embedded in a 
two--dimensional lattice.
(b) One--dimensional projection of the configuration of (a) along the
direction of the applied field identified by the sequence of relative 
projected coordinates $y = \{ 1, 0, 1, 1, -1, 1, 0 \}$. The vertical 
arrows represent the allowed moves for the reptons.
}
\label{FIG01}
\end{figure}

We consider here a $d$-dimensional version of the RD model on a hypercubic 
lattice (see Fig.\ref{FIG01}(a)). The polymer is divided in $N$ segments, 
or reptons, of the size of the order of the persistence length and each 
lattice site can accommodate an unlimited number of them. 
It is convenient to introduce a small driving external field $\varepsilon$, 
applied in a direction tilted by $45^\circ$ degrees with respect of the axes 
of the lattice: following previous work \cite{widom} we assign a rate $B =
\exp{(\varepsilon/2)}$ for moves of reptons in the direction of the field,
while moves in the opposite direction occur with a rate $B^{-1}$.
Here $\varepsilon$ is a dimensionless unit for the strength of the driving
field.  We focus here on the properties in the limiting regime of small 
$\varepsilon$, although the DMRG technique is not restricted to this regime.

The stationary state properties of the system can be derived from the
solution of the Master equation 
\begin{eqnarray}
\frac{d P(y,t)}{dt} &=& \sum_{y'} 
\left[ W(y|y') P(y',t) - W(y'|y) P(y,t) \right]
\nonumber \\
&=& \sum_{y'} H_{yy'} P(y',t)
\label{mastereq}
\end{eqnarray}
in the limit $t \to \infty$. Here $P(y,t)$ indicates the probability of 
finding the polymer in a configuration $y$ at time $t$ and $W(y'|y)$ is 
a transition rate per unit of time from a configuration $y$ to a configuration 
$y'$. The matrix $H$ contains both the gain and loss terms and is stochastic 
in the sense that the sum over all columns vanishes, as required from 
the conservation of probability.

Since the transition probabilities depend only on the projected
coordinate along the field direction the RD model becomes essentially
one dimensional (see Fig.\ref{FIG01}(b)). The relative coordinates
between neighboring reptons can assume only three values $y = \{-1, 0,
+1 \}$, therefore for a chain of $N$ reptons there are $3^{N-1}$
possible configurations. One should distinguish between moves for internal
and end reptons. In terms of the $y$ coordinates the moves are:
(a) Exchange of $0$'s and $1$'s for internal reptons, i.e. $\pm 1, 0
\leftrightarrow 0, \pm 1$, (b) end repton contractions $\pm 1
\rightarrow 0$ and (c) end repton stretchings $0 \rightarrow \pm 1$. 

The only effect of the dimensionality appears on the rates for the 
moves (c), which are $d B$ ($d B^{-1}$) for moves in the direction of 
(opposite to) the field, as the end repton can move to $d$ unoccupied new
sites (here dimension stands basically for the lattice coordination number
\cite{Rubin}).  Rates for the moves of type (a) and (b) are not 
affected by $d$. In general the dimensionality is believed to
have no influence on the dependence of the properties on N, as far as the
exponents of the asymptotic behavior are concerned. On a quantitave level,
as we shall see, the influence of $d$ is however large.
Higher dimensions stretch the polymer by giving a larger rate to the 
stretching processes (c). 
Not only integer values of $d$ are physically relevant, since the dimension 
enters in the RD model as the ratio between end--point stretching and 
end--point contraction. This allows to consider also values $d<1$. The limit 
of small $d$ corresponds to the case where the motions of type (c) are
suppressed. We will show that $d$  strongly influences the crossover
behavior and the appearance of effective exponents.  

In principle a direct numerical diagonalization of the matrix $H$ yields, for 
instance, the polymer drift velocity and the relaxation time \cite{widom}. 
In practice such calculations are restricted to polymers of rather short 
lengths ($N \leq 20$ \cite{widom,barkema97}) as the dimensionality of the 
configurational space grows exponentially with the number of reptons $N$. 
This problem can be overcome by the DMRG technique \cite{DMRGbook} that we 
apply here to the RD model. DMRG has been used for quite some time now and 
it is known for its accuracy and for the possibility of treating large 
systems using an efficient truncated basis set. Although originally 
introduced for hermitian matrices \cite{Whit92}, the DMRG method has also 
been applied to a series of non-hermitian problems \cite{nonherm}.
In the latter case the results are known to be less accurate than in the
hermitian case, however since we are primarily interested on the small
field limit the matrices to be diagonalized are only weakly
non-hermitian, therefore DMRG is expected to perform well in this case.

{\it Renewal time} -- We set first $\varepsilon = 0$. In this limit the 
matrix $H$ is symmetric since moves in the direction and opposite to the field 
are equally probable. The longest relaxation time $\tau$, also known as polymer
renewal time, is a quantity also accessible to experiments since it is related 
to viscosity. 
Theoretical arguments predict a scaling behavior as function of the chain 
length of the type $\tau \sim N^z$, with $z=3$ \cite{deGennes}. Various 
experiments yielded numerical values of the exponents systematically higher 
$z \approx 3.2 - 3.4$ (see, for instance, \cite{deGennes,doi,Rubin,crossover} 
and references therein). This apparent disagreement has generated some 
discussions in the past years and quite some effort has been devoted to 
reconcile theory with experiments. Doi \cite{crossover} argued that the 
discrepancy is due to finite size effects and proposed the following 
expression
\begin{equation}
\tau \sim N^3 \left[ 1 - \sqrt{\frac{N_0}{N}} \,\, \right] ^2,
\label{doi}
\end{equation}
with $N_0$ a characteristic length such that only for $N \gg N_0$ the right 
asymptotic behavior can be observed. Rubinstein \cite{Rubin} showed that a 
numerical calculation in the RD model (for chains up to $N \approx 100$ 
reptons) yielded $\tau \sim N^{3.4}$, however the asymptotic behavior $N^3$ 
was not observed.

\begin{figure}[b]
\centerline{
\psfig{file=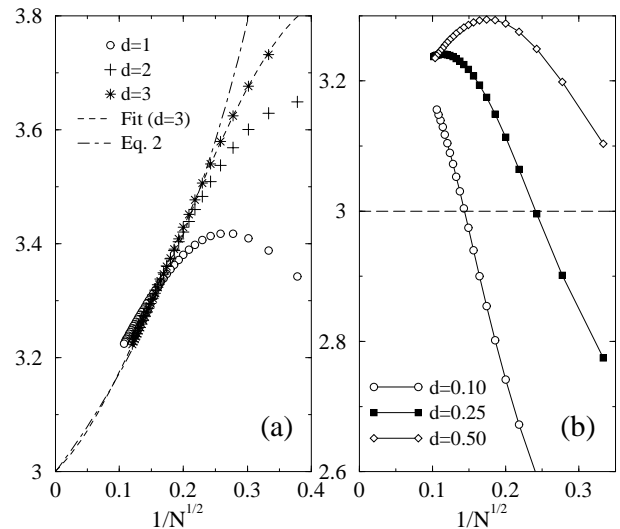,height=7.2cm}}
\vskip 0.2truecm
\caption{(a) Plot of $z_N$ as function of $N^{-1/2}$ for $d=1,2,3$; the 
dashed line is a cubic fit in powers of $N^{-1/2}$ for the $d=3$ case and 
the dotted-dashed line is obtained from Eq. \ref{doi}. 
(b) As in (a) for $d=0.10$, $d=0.25$ and $d=0.50$.
Here and in all the other figures error bars, unless explicitely shown, 
are smaller than symbol sizes.
}
\label{FIG02}
\end{figure}   

The renewal time is the inverse gap of the matrix
$H$, a quantity which is easily accessible in the DMRG calculation.
To estimate the exponent $z$ we considered:
\begin{equation}
z_N = \frac{\ln \tau_{N+1} - \ln \tau_{N-1}}{\ln (N+1) - \ln (N-1)}
\label{zN}
\end{equation}
which converges to $z$ in the limit $N \to \infty$.

In Fig. \ref{FIG02}(a) we show $z_N$ for $d = 1$, $2$, $3$ plotted as a 
function of $1/\sqrt{N}$ since this type of correction-to-scaling term
is predicted by  Eq.\ref{doi}. 
Using a cubic fit in powers of $1/\sqrt{N}$ we find $z = 3.00(2)$ for $d=3$, 
$z = 3.002(4)$ for $d=2$ and $z = 2.99(1)$ for $d=1$, in agreement with the 
theoretical arguments leading to a renewal time scaling as $\tau \sim N^3$ 
\cite{deGennes}. Figure \ref{FIG02}(a) shows also the fitting curve for $d=3$
(dashed) and the prediction from Doi's theory obtained by substituting 
Eq.(\ref{doi}) into Eq.(\ref{zN}) (dot-dashed). The latter compares very
well with our data in the region of large N, but fails to reproduce the 
change of curvature in $z_N$, which is characteristic for the DMRG results.

While in the case $d=2$, $3$ the asymptotic value is approached 
monotonically from above, for $d=1$ $z_N$ shows a non-monotonic
behavior. In a plot of $\ln \tau_N$ vs. $\ln N$ the maximum of $z_N$
corresponds to an inflection point, where curvature is absent.
Therefore in a range of lengths around the maximum the numerical data
would be fitted rather well by an effective exponent $z_{\rm eff} = 
\max_N z_N$. This is particularly true in the case of data affected by
small, but non-negligible error bars, as it happens in experiments or in
computer simulations. A monotonic behavior is less harmful since in a
$\ln \tau_N$ vs. $\ln N$ plot one should be able to distinguish always
a non-vanishing curvature, although it may happen that the data are well
fitted with effective powers in a range of values of $N$ also in that case.
We refer to the non-monotonic behavior as to a crossover effect. To
investigate it more into details we considered also $d < 1$ (see Fig.
\ref{FIG02}(b)). For $d=0.5$ and $d=0.25$ again the data show a 
non-monotonic behavior in $N$ and the turning point shifts to larger $N$, 
i.e. crossover effects are more pronounced.
For $d=0.1$, the smallest value considered here, and for the range of
$N$ investigated, $z_N$ increases monotonically. Assuming that the asymptotic
exponents do not depend on $d$, the turning point will be reached with much 
longer polymers than those considered here.
It is also important to point out that the values for the effective
exponent that we estimate from our data $z_{\rm eff} \approx 3.2-3.4$ 
are exactly in the range of values which are found experimentally in 
measures of viscosity \cite{doi}.
Therefore our results strongly support the idea that the differences
between theory and experiments are due to a crossover effect, as pointed
out by some authors in the past \cite{crossover,Rubin}. The advantage of 
the DMRG calculations is here that they provide clear information of both 
the asymptotic and non-asymptotic regions. Moreover, in the case of $d$ small,
we find a range of lengths for which the renewal time scales with a
local exponent smaller than 3 (see Fig. \ref{FIG02}(b)). The crossover to
the asymptotic regime where $z=3$ is approached from above (as in
Eq. \ref{doi}) occurs only for extremely long polymers. We are not 
aware of any experiments showing this type of behavior; possible candidates
could be concentrated polymer solutions with bulky endgroups in such a way
that endpoint strechings are suppressed.

{\it Diffusion constant} --
Crossover behavior appears also in other quantities, as for instance in the 
scaling of the diffusion constant $D(N)$ as function of the polymer length 
$N$. We calculated $D(N)$ by applying a small field $\varepsilon$ and using 
the Nernst--Einstein relation \cite{NE}
\begin{equation}
D = \lim_{\varepsilon \to 0} \frac{v}{N \varepsilon}.
\end{equation}
For the scaling behavior of $D(N)$ one expects
\begin{equation}
D(N) = \frac{1}{A N^2} \left( 1 + \frac{B}{N^\gamma} \right)
\label{diffusionc}
\end{equation}
where the leading term $N^{-2}$ is by now well-understood 
\cite{deGennes,doi,Rubin,barkema,periodic,prah,krenzlin} and it is 
considered to be experimentally verified \cite{rubber}. 
In the RD model also the prefactor happens to be known exactly 
\cite{periodic,prah}: $A = 2d +1$.
The next to leading order term has been investigated as well. By relating 
the diffusion constant to the renewal time it was predicted \cite{deutsch} 
that the correction term would be anomalous, i.e. $\gamma = 1/2$. This 
prediction is also supported by other theoretical arguments \cite{prah}. 
On the other side accurate Monte Carlo simulation results, done for $d=1$, 
could be best fitted with a power $\gamma \approx 2/3$ both for the RD
model \cite{barkema} and also for another model of polymer reptation
\cite{krenzlin}. This issue is still unresolved. An exponent $2/3$ is 
somewhat surprising since, as also seen above for the gap one naturally 
expects $N^{-1/2}$ corrections.

\begin{figure}[b]
\centerline{
\psfig{file=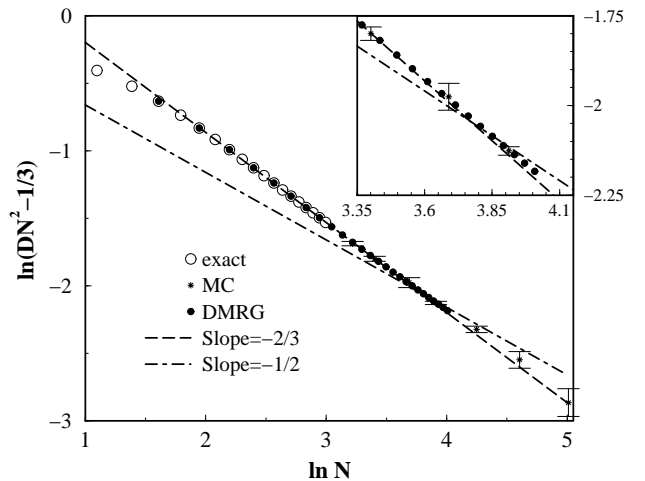,height=7.2cm}}
\vskip 0.2truecm
\caption{
Log-log plot of $D N^2 -1/3$ vs. $N$ for $d=1$. The DMRG data are in good
agreement with the exact ones and with the Monte Carlo results. Inset: blow up
of the region for large $N$ where the deviation of the DMRG data from the
slope $2/3$ starts being noticeable.
}
\label{FIG03}
\end{figure}   

In Fig. \ref{FIG03} we show a plot of $g_N = D(N) N^2 - 1/A$ as function of 
$N$ on a log-log scale for $d=1$. The figure shows results from exact 
diagonalization for small lattices, Monte Carlo simulations \cite{barkema97} 
and DMRG results which we extended up to $N=57$.
DMRG data are in good agreement with those obtained from other 
methods and follow rather nicely a slope $-2/3$ in the plot. Only a very close
inspection of the region of large-$N$ systems reveals (see inset)
that this slope is not the correct asymptotic one. Similar to the gap, 
this can be best seen from the discrete derivative
\begin{equation}
\gamma_N =  - \frac{\ln g_{N+1} - \ln g_{N-1}}{\ln (N+1) - \ln (N-1)}.
\end{equation}

\begin{figure}[b]
\centerline{
\psfig{file=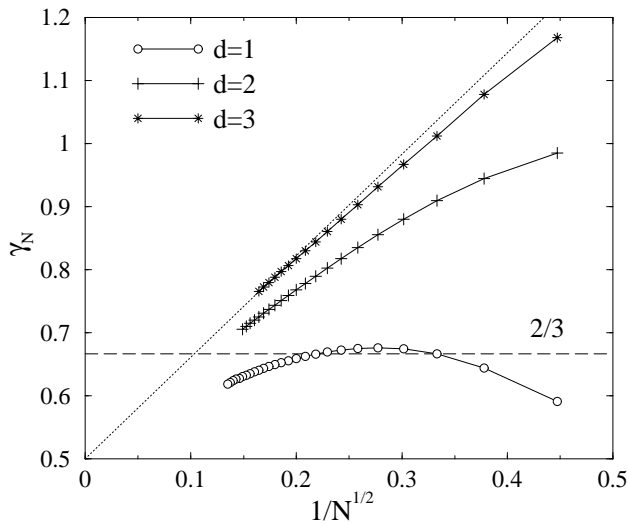,height=7.2cm}}
\vskip 0.2truecm
\caption{
Plot of the effective exponent $\gamma_N$ as function of $1/\sqrt{N}$
for $d=1$, $2$ and $3$. 
The dashed line is the value of the correction exponent conjectured 
on the basis of Monte Carlo simulations for $d=1$. The dotted line 
is a guide for the eye.
}
\label{FIG04}
\end{figure}

A plot of $\gamma_N$ vs. $1/\sqrt{N}$ is shown in Fig. \ref{FIG04} for
$d=1$, $2$ and $3$.
As for the gap, we note a non-monotonic behavior for $d=1$, where the maximum 
of $\gamma_N$ is found at about $0.68$, i.e. very close to the power
$\gamma=2/3$. However for sufficiently large $N$, $\gamma_N$ clearly
deviates from $2/3$ to smaller values. For $d=2$ and $3$, $\gamma_N$ is 
monotonic in $N$. 

Like for the renewal time exponent $z_N$, we fit $\gamma_N$ with a
cubic curve containing powers of $1/\sqrt{N}$
\begin{equation}
\gamma_N = \gamma + \sum_{i=1}^3 \frac{\alpha_i}{N^{i/2}} ,
\end{equation}
where $\gamma$ and $\alpha_i$ are fitting parameters.
Extrapolations yield $\gamma = 0.51(1)$ ($d=3$), $\gamma = 0.51(1)$ ($d=2$) 
and $\gamma = 0.48(3)$ ($d=1$). These results strongly support a correction 
term with $\gamma = 1/2$.

In conclusion, the above results show that DRMG is powerful technique to 
investigate the properties of a reptating polymer. In this Letter we have 
restricted ourselves to two aspects: the renewal time and the diffusion 
coefficient, since these quantities have been mostly debated in the past. 
The DMRG calculations also yield a host of detailed information about the 
structure of the reptating polymer. In a forthcoming publication we will 
present data on the drift velocity for larger values of $\varepsilon$ and on 
structural properties of the chain such as the averages $\langle y_i 
\rangle$ and $\langle y_i y_{i+1} \rangle$, where $i$ labels the position 
along the chain.
Here we have shown that the large finite size corrections, 
characteristic for the reptation process, manifest themselves as effective 
exponents for the asymptotic behavior of the renewal time and the diffusion 
coefficient.
The accurate values for a large set of lengths $N$ make it possible to
determine the corrections to scaling with great precision. In this way we 
find a natural reconciliation of the fairly strong theoretical arguments for 
a renewal time exponent $z=3$ and the equally pertinent experimental findings 
of values around $z \approx 3.2-3.4$. Our analysis also shows that log--log 
plots to determine exponents are hazardous when such large corrections to 
scaling are present. In particular when the effective   
exponent shows a stationary point as function of $N$, it shows up as an
inflection point which can be easily mistaken for the asymptotic region
in case of insufficient data. On the other hand the method of differential 
exponents is a very accurate indicator of the asymptotic behavior. However 
differential exponents require precision data for many values of the length 
$N$, which neither in Monte Carlo simulations nor in experimental measurements
are feasible.

We are grateful to G.T. Barkema for helpful discussions.

\end{multicols} 

\end{document}